# The Multidimensional Study of Viral Campaigns as Branching Processes

Jarosław Jankowski[1], Radosław Michalski[2], Przemysław Kazienko[2]

[1] Faculty of Computer Science, West Pomeranian University of Technology, Szczecin, Poland
`jjankowski@wi.zut.edu.pl`
[2] Institute of Informatics, Wrocław University of Technology, Wrocław, Poland
`radoslaw.michalski@pwr.wroc.pl, kazienko@pwr.wroc.pl`

**Abstract.** Viral campaigns on the Internet may follow variety of models, depending on the content, incentives, personal attitudes of sender and recipient to the content and other factors. Due to the fact that the knowledge of the campaign specifics is essential for the campaign managers, researchers are constantly evaluating models and real-world data. The goal of this article is to present the new knowledge obtained from studying two viral campaigns that took place in a virtual world which followed the branching process. The results show that it is possible to reduce the time needed to estimate the model parameters of the campaign and, moreover, some important aspects of time-generations relationship are presented.

## 1    Introduction

There are variety of models to describe the diffusion of information: starting from epidemic models [24], [2], through the Bass model [5], adoption model [8], [29] and variety of models combining network and user properties [14], [23], [7], [3]. Another model, the super-diffusion [26], [27], [31], may be called the fastest diffusion across all the previously mentioned. A good comparison of information diffusion models may be found in [28] and more recent research regarding new models is presented in [16].

Yet another model, the branching process [21] is since recently used for analysing diffusion of information extending the ability to describe how the information may flow through the social network [22], [25], [17], [18]. Even some suggestions may be found that the branching model is more adequate to describe the information diffusion, especially while compared to the disease spreading models [18]. However, it is not a question of preference while fitting real-world data to a particular model – the goodness of fit decides which model a particular viral campaign follows. The outcomes of having a model and its parameters discovered are that one may benefit from this general model properties and implications which make the analysis of the cam-paign easier. Generally, it is preferred to know which model a particular campaign follows while it happens, because in that case it would be easier to predict further behaviour of the information diffusion process. However, in that case, the branching process introduces some limitations described in Section 2.

This paper focuses on analysing two campaigns conducted in virtual world environment which followed the branching process model to present valuable outcomes in terms of reducing the need for having all the campaign data to calculate branching process parameters. To extend this result in terms of time aspects, authors decided to deepen the analysis towards the relationship between the time and generations in this model which led to better knowledge about how particular branches of the model are being developed in time, what may even allow to predict on-going campaigns results.

The structure of this paper is as follows. The next section of this paper describes the problem, which is followed by the related work analysis. Section 4 presents the experiment setup and the description of the analysed campaigns. Experimental studies results are presented in Section 5 with the conclusions and future work directions presented in Section 6.

## 2   Problem Description

The branching process in terms of information diffusion may be basically described as a process where an individual may spread the information to a number of consequents. Starting from a number of seeds understood as the first generation the information is forwarded towards next generations, creating a tree of information traversal. The information diffusion ends when there will be no further infections, that means reaching the all of the susceptible users. The nature of the branching process, especially the fact that it is not based on time but on generations, makes the whole process a bit harder to interpret on a time basis, because the number of users infected in a particular generation changes over time. And as the basic equations of the branching process are calculating the number of infected in the next generation basing on the previous one, the chance to estimate the parameters while the campaign is on-going is very weak, unless there exists a certainty that the number of infected users in previous generations would not change, which is rather unlikely in real campaigns. That results in constant underestimation of the overall number of infected users.

Basing on the above limitations of the branching model, authors of this paper decided to examine whether is there any chance to estimate the total number of infected users in the viral campaign while using only a partial information of all infections. And in that case the partial information means the complete information about only part of consecutive generations starting from the first one. So, to describe the problem in more formal way, the task is to estimate the $p$, $N$ and $\lambda$ parameters for the model [22] by using real-world data in the way that only the complete information about particular number of generations is available beginning with the first generation.

The question arises why would one benefit from such an approach if it means that the analysis would be performed *a posteriori*? To name only the one major argument, the task to estimate the campaign parameters is to find a model which fits all the branching process generations as good as it is possible. So, in that case, it is necessary to find in a three-dimensional space a set of parameters which minimize the error of fit across all generations. If the approach proposed by the authors succeeds, the calculation time needed for estimating the parameters will reduce still providing good estimation. However, as it is described in Section 4, authors decided also to evaluate how are particular generations changing over time. And if it will be seen that all the generations needed for the model parameters estimation stabilize before the campaign ends, it will lead to the conclusion that the parameter estimation may be performed earlier as well.

So, despite these limitations, why is the branching process becoming more popular in modelling the diffusion of information? As described in [22], most of the models base on aggregated information about the total number of infections. In that case the advantage of the branching process is that its approach allows to analyse the epidemics on different level focused on individual reproduction rate, what may lead to extending the knowledge about the diffusion of information.

## 3   Related Work

The concept of fitting real life data to a particular model plays an important role in statistics. Goodness of fit tests are used to fit variety of data to the existing models, and social network analysis also often reaches for those tests. For instance, real life social networks are to be fitted to models [15], variety of analyses are performed to check how particular social network properties fit power law distribution [11], not mentioning on new model generation and fitting [30]. The benefit of having a particular process fit to the model is that in that case it is easier to generalize the observed data and predict the future behaviour of the information diffusion.

As it was stated in the previous Section, the experiments conducted by authors of this paper are regarding finding the optimal set of parameters for modelling the branching process in terms of limited knowledge about the branching process. In that case authors wanted to find the answer on the question whether is it possible to estimate the parameters of the model by having only the information about a few generations. Most of the literature regarding estimation of the branching processes is related to supercritical processes [12], [13], however those approaches were more related to obtaining the distribution of offspring probability. In [22] authors analysed the possibility of use the branching process to model the diffusion of information and decided to estimate the model parameters in discrete time what differs from the approach presented in this paper. The other type of analysis authors of this paper perform is the study of the relationship between generations and time. An interesting case of continuous time branching processes was described in [20], where next generations were strongly related to the time due to biological reasons. Another work on the problem of time-generation relationship in the branching process with regards to epidemics which also states that if the transitions are population dependent, the long-term prediction of these processes is an open problem is [19].

However, authors of this paper were unable to find similar to the presented approach in terms of modelling viral campaign by using consecutive generations and a single set of parameters, the deepened analysis of time-generations relationship was also not studied extensively by others.

## 4  Experimental Setup

Authors of this paper analysed two real-world campaigns from a virtual world environment with the goal to estimate the branching process parameters. However, the basic intuition in the branching process is that in every generation the model parameters will change due to increasing or decreasing interest in the campaign what may harden the task of predicting the final spread of the campaign. Authors decided to omit this problem by trying to estimate only a single set of parameters for the whole campaign.

The experiment setup was as follows: for the whole dataset number of infections in every generation was calculated. Next, starting from only the first generation branching model parameters were calculated and evaluated with the real data in terms of MSE errors of overall campaign reach and the cumulative MSE error calculated as the difference of real data reach and estimated reach for every generation. Next, this procedure was repeated for a model parameters estimated by using two consecutive generations starting from the first one, three generations and so on until the set of generations was equal to the number of generations in real-data. This procedure allowed to analyse how well the model built by using less number of generations is able to estimate the overall campaign reach and behaviour.

As it was already described in the Section 2, due to the changing number of infections for each generation in time, the proposed approach still requires to have the whole dataset to be applied. However, if the approach succeeds only a limited number of generations would be required to calculate the model parameters what decreases the overall processing time. But to get additional knowledge about the overall behaviour of the branching process, authors decided to extend the study by analysing how particular generations change over time. In that case if the experiment results will prove that the number of generations required to adequately model the data stabilize before the campaign ends, the additional outcome may be the ability to predict the campaign behaviour at earlier stage (on-going). So the second part of the research was devoted to analyse the time-generations relationship.

During the research, there were data from two viral actions with different characteristics from social platforms working in a form of virtual world. In both actions, users were spreading virtual goods like avatars using viral mechanism to their friends. The first viral action denoted as $V_1$ was based on sending gifts to friends and the senders' motivation to spread those gifts was not incentivized. The second action, denoted as $V_2$, was based on incentives and competition among users to spread visual elements of avatars among their friends.

The analysis of number of infections in time gives the knowledge about the dynamics of the campaign, however it is not delivering information about the structure of infections. Both campaigns had different specifics and by using aggregated models based on time dimension only, there is an additional analysis to show structures of infections needed. Next, both campaigns were analysed by using the proposed approach based on generations and parameters describing their characteristics. A generation is defined in terms of viral marketing as a number of transmissions required to reach a member along a chain of communication initiated by a single seed [4]. The approach based on generations can capture structures of transmissions which is not possible by the cumulative analysis based only on infections over the time. In the earlier research, three main parameters from epidemic theory were used for modelling the characteristics of viruses spreading which can be applicable to viral marketing campaigns as well [21]. Contagion parameter denoted as $p$ describes the probability of transferring viral message by an infective. Epidemic intensity $\lambda$ represents the number of customers reached. Epidemic threshold parameter $ETP$ defined as $p*\lambda$ describes the progression of epidemics. Becker defined relation between characteristics of campaign and epidemic threshold parameter as sub-critical ($ETP < 1$), super-critical ($ETP > 1$) and critical ($ETP = 1$) [6]. D. B. Stewart et al. presented conceptual framework for viral marketing [25]. The mathematical model presented by the authors for viral campaigns modelling was based on deterministic model discussed by J.C. Frauenthal [9], used by R.M. Anderson and R.M. Mayfor modelling epidemic [1], extended later by G. Fulford et al. [10].

## 5  Results

The empirical research was conducted by using the above described approach. The main goal was to conduct extended analysis of viral campaign using the approach based on branching processes and verify the ability to predict the viral campaign model by analysing two real datasets. In Figure 1 and Figure 2 the cumulative number of infections in the analysed time period (days) for both campaigns is presented.

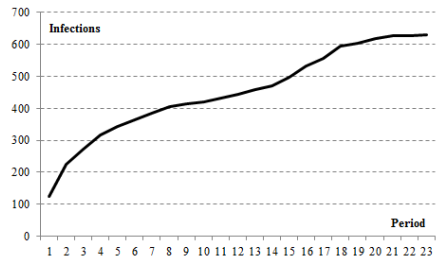   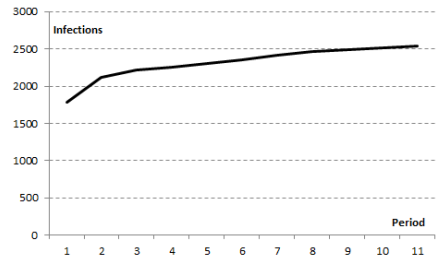

**Fig. 1.** Cumulative number of infections - $V_1$     **Fig. 2.** Cumulative number of infections - $V_2$

Following the earlier approaches, the analyses of two viral campaigns were performed and $p$, $\lambda$ and $ETP$ parameters were computed and showed in Table 1 for both of campaigns.

**Table 1.** Number of infections and parameters of analysed campaigns

|       | G  | Infected | Cumulative | Decisions | Infections sent | p      | λ       | ETP     |
|-------|----|----------|------------|-----------|-----------------|--------|---------|---------|
| $V_1$ | 1  | 1        | 1          | 1         | 11              | 1.0000 | 11.0000 | 11.0000 |
|       | 2  | 11       | 12         | 10        | 49              | 0.9091 | 4.9000  | 4.4545  |
|       | 3  | 49       | 61         | 26        | 106             | 0.5306 | 4.0769  | 2.1633  |
|       | 4  | 106      | 167        | 42        | 123             | 0.3962 | 2.9286  | 1.1604  |
|       | 5  | 123      | 290        | 41        | 90              | 0.3333 | 2.1951  | 0.7317  |
|       | 6  | 90       | 380        | 33        | 79              | 0.3667 | 2.3939  | 0.8778  |
|       | 7  | 79       | 459        | 20        | 41              | 0.2532 | 2.0500  | 0.5190  |
|       | 8  | 41       | 500        | 11        | 43              | 0.2683 | 3.9091  | 1.0488  |
|       | 9  | 43       | 543        | 12        | 40              | 0.2791 | 3.3333  | 0.9302  |
|       | 10 | 40       | 583        | 14        | 38              | 0.3500 | 2.7143  | 0.9500  |
|       | 11 | 38       | 621        | 7         | 13              | 0.1842 | 1.8571  | 0.3421  |
|       | 12 | 13       | 634        | 3         | 4               | 0.2308 | 1.3333  | 0.3077  |
|       | 13 | 4        | 638        | 1         | 1               | 0.2500 | 1.0000  | 0.2500  |
|       | 14 | 1        | 639        | 0         | 0               | 0.0000 | 0.0000  | 0.0000  |
| $V_2$ | 1  | 9        | 9          | 8         | 187             | 0.8889 | 23.3750 | 20.7778 |
|       | 2  | 187      | 196        | 52        | 552             | 0.2781 | 10.6154 | 2.9519  |
|       | 3  | 552      | 748        | 115       | 782             | 0.2083 | 6.8000  | 1.4167  |
|       | 4  | 782      | 1530       | 105       | 450             | 0.1343 | 4.2857  | 0.5754  |
|       | 5  | 450      | 1980       | 55        | 251             | 0.1222 | 4.5636  | 0.5578  |
|       | 6  | 251      | 2231       | 32        | 137             | 0.1275 | 4.2813  | 0.5458  |
|       | 7  | 137      | 2368       | 18        | 47              | 0.1314 | 2.6111  | 0.3431  |
|       | 8  | 47       | 2415       | 5         | 27              | 0.1064 | 5.4000  | 0.5745  |
|       | 9  | 27       | 2442       | 4         | 51              | 0.1481 | 12.7500 | 1.8889  |
|       | 10 | 51       | 2493       | 4         | 6               | 0.0784 | 1.5000  | 0.1176  |
|       | 11 | 6        | 2499       | 3         | 4               | 0.5000 | 1.3333  | 0.6667  |
|       | 12 | 4        | 2503       | 2         | 0               | 0.5000 | 0.0000  | 0.0000  |

As the above table shows, the epidemic parameters are changing over the time and describing campaign with a single set of parameters may be a challenging task. Due to these changes, it is difficult to predict the next stages of the campaign by using data from earlier periods. In the analysed campaign $V_1$ *ETP* for the first generation was equal 11 while in the second generation it was only 40.49% of earlier value being reduced to 4.4545. After few generations of decreasing, it went up to 1.0488 at generation number eight. For campaign $V_2$, even bigger changes were identified (possibly because of the incentives), especially between the first generation with *ETP*=20.78 reduced to 2.95 in the second generation. In the next stage of research, the analysis of change of the parameters was performed for both of campaigns. Campaign $V_1$ was identified as super-critical during generations $G_1$, $G_2$, $G_3$, $G_4$ and $G_8$ while $V_2$ can be treated as super-critical for generations $G_1$, $G_2$, $G_3$ and $G_9$. An interesting result is that characteristics of the second campaign show that incentives were not effective to increase number of generations characterized as super-critical according to *ETP*, so the campaign reach in terms of number of generations was similar. The other problem identified during the analysis is related to changes of data for each generation with time periods, as it was described in Section 2. If the analysis is performed in the first period, after the second period the results from earlier analysis are useless in terms of prediction, because of changes in all existing generations. Computations must be performed on the whole data and cannot be based on incremental approach. Changes in the number of infections for campaign $V_1$ are showed in the Table 2 for the first ten days of campaign. In the first period of time (the first day of the campaign), 19,56% of all infections where registered within ten generations. During the second day no additional infections occurred for $G_1$ and $G_2$ but additional infections are found for generations $G_3$-$G_{10}$. At this period, the first two infections are registered for generation $G_{11}$. Generations $G_{12}$ and $G_{13}$ are not built until period $P_5$ and $P_6$.

**Table 2.** Changes of number of infections in generations over the time

| G | Campaign period (in days) | | | | | | | | | |
|---|---|---|---|---|---|---|---|---|---|---|
|   | 1 | 2 | 3 | 4 | 5 | 6 | 7 | 8 | 9 | 10 |
| 1  | 1  | 0  | 0  | 0  | 0 | 0 | 0 | 0 | 0 | 0 |
| 2  | 10 | 0  | 0  | 0  | 0 | 0 | 0 | 0 | 0 | 0 |
| 3  | 29 | 6  | 5  | 1  | 0 | 2 | 3 | 0 | 1 | 0 |
| 4  | 40 | 25 | 13 | 8  | 0 | 0 | 3 | 0 | 2 | 0 |
| 5  | 22 | 31 | 12 | 7  | 3 | 1 | 1 | 3 | 3 | 2 |
| 6  | 14 | 7  | 11 | 3  | 5 | 7 | 2 | 0 | 0 | 1 |
| 7  | 3  | 12 | 2  | 0  | 3 | 4 | 0 | 4 | 3 | 3 |
| 8  | 2  | 7  | 0  | 0  | 4 | 0 | 1 | 3 | 1 | 0 |
| 9  | 2  | 6  | 3  | 5  | 4 | 2 | 5 | 2 | 0 | 0 |
| 10 | 2  | 4  | 1  | 4  | 5 | 2 | 3 | 3 | 1 | 0 |
| 11 | 0  | 2  | 1  | 15 | 3 | 0 | 1 | 3 | 0 | 0 |
| 12 | 0  | 0  | 0  | 0  | 1 | 1 | 1 | 2 | 0 | 0 |
| 13 | 0  | 0  | 0  | 0  | 0 | 1 | 0 | 0 | 0 | 0 |
| 14 | 0  | 0  | 0  | 0  | 0 | 0 | 0 | 0 | 0 | 0 |
|    | 19.56% | 15.65% | 7.51% | 6.73% | 4.38% | 3.13% | 3.13% | 3.13% | 1.72% | 0.94% |

The cumulative number of infections for chosen generations is presented in the Figure 3. It shows that the dynamics of increase of infections in generations is changing over the time. In the example presented, for generation G4 during the first four periods the increase of number of infections was observed and during the next period it stabilizes. The situation changes slightly at period 16 when the next increase is observed. For earlier generations situation is more stabilized. For generations G5 and G6, growth is observed until the 21st period, but highest changes were identified in periods 1-5.

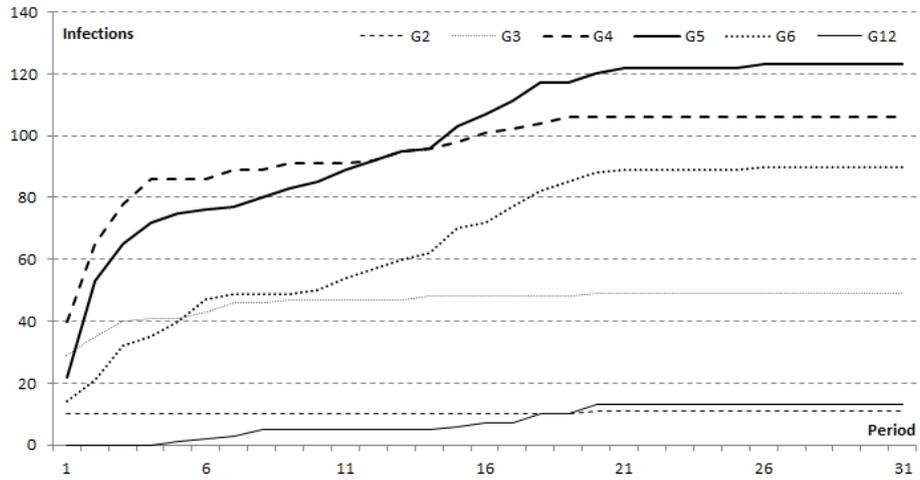

**Fig. 3.** Cumulative number of infections for selected generations in campaign $V_1$

In the next step, for both campaigns the dynamics of new generations creation was analysed and the results are presented in Figure 4. The figure presents the moment of the first occurrence of the particular generation. Even though the whole data set for campaign $V_1$ is based on 31 days and $V_2$ on 11 days, the results showed that during the first two hours of campaign ten generations were created for campaign $V_1$ and seven for campaign $V_2$. Dynamics of creation of new generations was higher for campaign $V_2$ with incentives and during first twenty five minutes six generations were created while for first campaigns six generations were created during sixty minutes. It shows that campaigns go very fast in depth (vertically) and after that they start to grow horizontally.

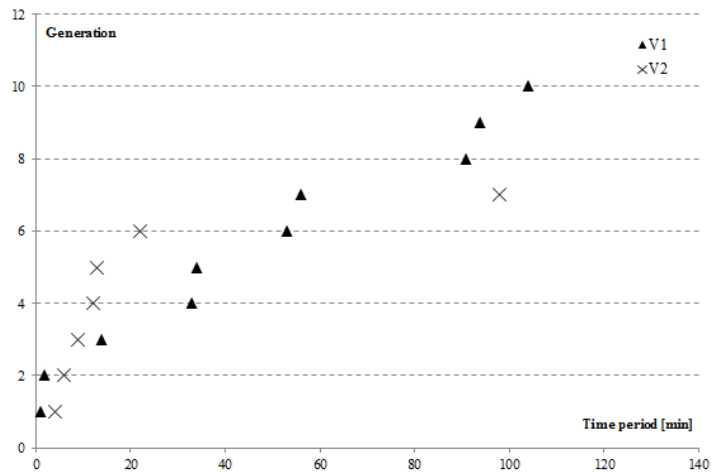

**Fig. 4.** The dynamics of generations' creation for campaigns $V_1$ and $V_2$

The presented analysis show how the generations-based approach can be used to analyse characteristics of viral campaign. Volatile data and change over the time make it difficult to build a prediction model using data from past periods due to changes in generations. The results presented show that estimation of parameters for viral campaigns based on generations approach using contagion parameter and epidemic intensity needs computations with every time period after data increase in all generations. Apart from this it is difficult to estimate parameters because of changes at all generation. The most convenient way would be describing campaign with a single set of $p$ and $\lambda$ parameters for the whole campaign. It would be easy under the assumption that those parameters are stable during whole campaign. In the next stage experimental results of the proposed method are presented for searching the best fitting model and estimating the campaign reach over consecutive generations. The statistics for the campaigns were used as a reference and the main goal was to estimate campaign reach and to generate a model describing campaign performance with the minimal possible set of generations. For each stage of the campaign, starting from the first generation a branching model is built and parameters are adjusted to find the best fit. Estimations were computed for both campaigns and all generation sets, starting from a set consisting of one generation towards the set combined of all generations, which, in fact, was the reference model. Results for campaign $V_1$ are showed in Figure 5.

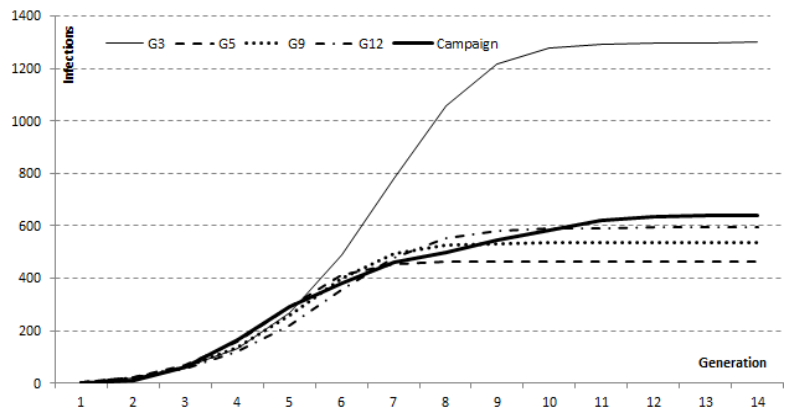

**Fig. 5.** Estimations for campaign $V_1$

The estimation computed at generation three resulted in estimation of campaign reach at level of 1299.10 while real campaign reach was 639. Computations performed after the fifth generation resulted in an estimated reach at level 461.97 with 27.70% reach error comparing to the full campaign dataset. Estimation after generation $G_9$ resulted in 16.35% reach error and after $G_{12}$ - 7.23%. The results for the campaign showed that at the sixth generation $G_6$, it was possible to estimate the model acceptable in terms of minimizing the MSE over all generations. As it may be seen, the model parameters gave the opportunity to predict the reach of the campaign accu-rately. Similarly, the quality of estimation was computed by using mean square error and there was a model selected giving the lowest error for all generations evaluated for campaign $V_2$. Computations after third generation predicted a total of 3867 reach, for the sixth predicted the reach was 2000 and for ninth while the observed campaign reach was 2536. In Table 3, the detailed errors computed for estimated model parameters are showed and the differences between reach predicted from model and for campaign $V_1$.

**Table 3.** Mean square errors and reach errors for campaign $V_1$

| G | Period MSE | Campaign MSE | Estimated reach | Reach error | Reach error [%] |
|---|---|---|---|---|---|
| 1 | 0.00 | 207295.54 | 5.15 | 633.85 | 99.19% |
| 2 | 0.16 | 140916.39 | 142.38 | 496.62 | 77.72% |
| 3 | 26.68 | 223639.37 | 1299.10 | 660.10 | 103.30% |
| 4 | 47.97 | 146480.08 | 1103.70 | 464.70 | 72.72% |
| **5** | **56.66** | **10083.63** | **461.97** | **177.03** | **27.70%** |
| 6 | 142.19 | 10798.72 | 456.67 | 182.33 | 28.53% |
| 7 | 153.49 | 10798.72 | 456.67 | 182.33 | 28.53% |
| 8 | 1103.10 | 8900.25 | 475.36 | 163.64 | 25.61% |
| 9 | 454.36 | 3250.94 | 534.50 | 104.50 | 16.35% |
| 10 | 647.85 | 3250.94 | 534.50 | 104.50 | 16.35% |
| 11 | 1198.52 | 1363.36 | 592.79 | 46.21 | 7.23% |
| 12 | 1241.85 | 1363.36 | 592.79 | 46.21 | 7.23% |
| 13 | 1303.94 | 1363.36 | 592.79 | 46.21 | 7.23% |
| 14 | 1353.71 | 1353.71 | 604.17 | 34.83 | 5.45% |

The results in the Period MSE column show errors computed when comparing model chart for selected number of generations with real data. The column Campaign MSE shows error after comparing the model for all generations with the campaign data. Results for campaign $V_1$ show that it is possible to build the model of the campaign with acceptable reach error by using only data from five generations (35% of all generations). The reach error decrease for both of the campaigns is illustrated in Figure 6.

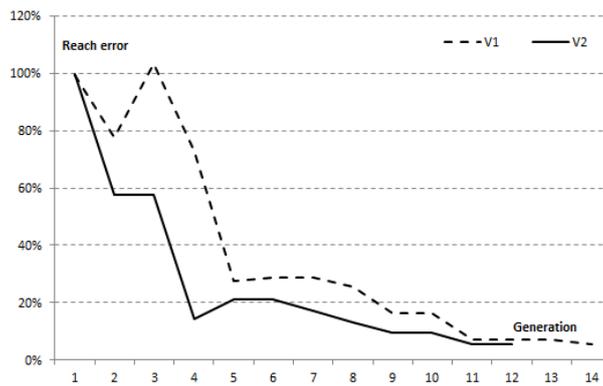

**Fig. 6.** The relationship of number of generations used for estimation and the reach error

## 6      Conclusions and Future Work

The presented research showed multidimensional approach to viral campaign analysis by using branching processes as a model and view on campaigns based on generations. By comparing to the time-based analysis, it is possible to catch some deepened characteristics and interesting results. Recently some research was focused on applications of branching processes in viral marketing but it was mainly focused on building models of campaigns on whole datasets and changes within generations were not discussed.

The research focused on technical analysis of the distribution of media without taking into account the social aspects. This approach was targeted to separate the components of generalized characteristic viral campaign from the social factors that may be unique to the analysed environment. This approach, however, does not exclude ability to use social network characteristics and take into account the distribution network and the attributes characterizing the participants in the campaign.

This analysis based on branching processes approach delivered information about different specifics of both campaigns. Generations give the possibility to analyse structure of infections and make it possible to observe dynamics on each level. Campaigns performance can be compared and the results obtained can be used to evaluate effectiveness and detect drop or increase in the campaign dynamics. It showed different dynamics of changes at generation level for both campaigns.

Apart from extended data analysis presented in this research, the method of building branching model with parameters based on best fit model not using data from all stages of campaign was also presented. The presented approach makes it possible to predict campaign reach without analysing campaign parameters for each generation. Results based on real campaigns showed that it was possible to estimate the campaign reach after fifth generation while whole dataset had fourteen generations. This approach is useful for situations when changes in the parameters make it difficult to describe the whole campaign with only two parameters which are stable for all stages like contagion and epidemic intensity. In terms of studying campaigns which are ongoing, the results from the proposed method are dependent on stabilization of infections in generations used for computations and the proposed approach may be used as a predicting one only after required generations will stabilize. That means generations should be included in the set used for the estimation of parameter when no dynamic growth is observed in number of infections. Apart from selecting the best fit model, the proposed approach can be used to build the knowledge base on campaign structures and instead of comparing the campaign to the model real campaigns data can be used as well.

Despite extending the scientific knowledge in the topic, this sort of knowledge may be found out as valuable for practitioners. Especially the time-generation analysis results show that at the very beginning we are able to see how deep the campaign will go – in that case the early knowledge about the possible reach of the campaign may give the campaign managers additional time to start the campaigns in different social network areas as well.

The presented research opens new research questions which may be explored further. For performance and quality of predictions a method can be developed to predict changes in generations in the next periods to detect a moment of time when stabilization is expected to include generation in the set for computations. Factors and characteristics of campaign affecting community exploration with generations and relations between the number of infections in each generation were also found to be an interesting area for the next research, as well as research based on simulation data which shows for what kind of networks' and parameters' models the proposed approach is suitable.

**Acknowledgments.** This work was partially supported by fellowship co-financed by the European Union within the European Social Fund, the Polish Ministry of Science and Higher Education, the research project 2010-13.